# Multi-level Chirality in Liquid Crystals Formed by Achiral Molecules

Mirosław Salamończyk[1,2], Nataša Vaupotič[3,4*], Damian Pociecha[1], Rebecca Walker[5], John M. D. Storey[5], Corrie T. Imrie[5], Cheng Wang[2], Chenhui Zhu[2], Ewa Gorecka[1*]

[1]University of Warsaw, Faculty of Chemistry, Żwirki i Wigury 101, 02-089 Warszawa, Poland.

[2]Lawrence Berkeley National Laboratory, Advanced Light Source, 1 Cyclotron Rd, Berkeley, CA 94720, USA.

[3]Department of Physics, Faculty of Natural Sciences and Mathematics, University of Maribor, Koroška 160, 2000 Maribor, Slovenia.

[4]Jozef Stefan Institute, Jamova 39, 1000 Ljubljana, Slovenia.

[5]Department of Chemistry, King's College, University of Aberdeen, Aberdeen AB24 3UE, UK.

*Correspondence to: natasa.vaupotic@um.si (N.V.), gorecka@chem.uw.edu.pl (E.G.).

**Abstract**: In many biological materials with a hierarchical structure there is an intriguing and unique mechanism responsible for the 'propagation' of order from the molecular to the nano- or micro-scale level. Here we present a much simpler molecular system built of achiral mesogenic dimeric molecules that shows a similar complexity with four levels of structural chirality (*i*) layer chirality, (*ii*) helicity of a basic 4-layer repeating unit, (*iii*) a helix with a pitch of several layers and (*iv*) mesoscopic helical filaments. As seen in many biological systems, there is a coupling between chirality at different levels. The structures were identified by a combination of hard and soft x-ray diffraction measurements, optical studies and theoretical modelling.

**One Sentence Summary**: Simple molecular system exhibits complex hierarchical structure with chirality coupling between different structural levels.

It is well known that in biological systems the molecular chirality of nucleic acids and sugars renders the formation of the helical meso-structure of DNA and proteins that finally leads to a lack of mirror symmetry of macro bio-objects. Hierarchical structures are also inherent to simpler molecular systems like liquid crystals. The presence of a chiral center in the structure of a mesogen can dramatically alter the liquid-crystalline phase behavior and therefore, material properties. If chiral molecules form a nematic or tilted smectic phase the neighboring molecules tend to twist with respect to each other and a long-range helix is formed (*1*). In some cases, the structure is further transformed into a more complex one, for example, into the cubic lattice of a blue phase (*2*) or a twist grain boundary (*TGB*) phase (*3*). Because chiral interactions are weak, the helical pitch is usually long and the repeating unit of these 3D structures is fairly large, usually hundreds of nanometers. In recent years, it became apparent that twisted structures are not unique to chiral molecules, with the discovery that although achiral, bent-core molecules or bent mesogenic dimers can form twisted structures with a surprisingly short pitch of just a few nanometers. The twist-bend nematic ($N_{TB}$) phase (*4–8*) is the simplest heliconical structure, and found for a number of odd-membered dimers and for some rigid bent-core molecules (*9, 10*). In the $N_{TB}$ phase the molecules precess along the helix axis, on average being inclined to the axis by an angle θ (tilt angle). A regular helical arrangements that are self-repeating in 3D have been proposed for cubic or tetragonal phases formed by multiple end chain molecules (*11–15*). On the other hand, there are still only a few examples of smectic systems formed by achiral molecules for which helical structures were suggested (*16–18*) and just one for which it was unambiguously proved (*19*).

In this letter, we report on two molecular systems in which the competing interactions lead to a complex, multi-hierarchical helical structure of a lamellar phase. The nonsymmetric dimer *D1* shows a helical, liquid-like smectic phase below the $N_{TB}$ phase and the nonsymmetric dimer *D2* forms a helical lamellar phase below an orthogonal smectic phase. The molecular structures of these materials and their phase transition temperatures are given in Fig. 1. The structures of the smectic phases were resolved by resonant x-ray scattering studies (RSoXS). In contrast to non-resonant x-ray diffraction which is sensitive only to the electron density modulations, RSoXS also detects the spatial variation of the orientation of molecules (*20*). In our studies we use RSoXS at the carbon absorption K-edge; this experimental method requires a very low energy beam (~284 eV), but is universal, as carbon atoms are present in every organic molecule. The method based on spectroscopic contrast has recently been successfully applied to study a phase separation in block copolymers (*21*) and molecular orientation in solar cells (*22*). It also revealed the morphology of helical nanofilaments (*B4* phase) made of bent-core mesogens (*23*) and the twisted structure of liquid crystalline phases having no positional order (*24–26*). Apart from the RSoXS measurements we also performed careful non-resonant x-ray diffraction studies and investigated the optical properties of the materials. Experiments were complemented by theoretical modelling. Details of the experimental methods and theoretical calculations are given in the Supplementary material (SM).

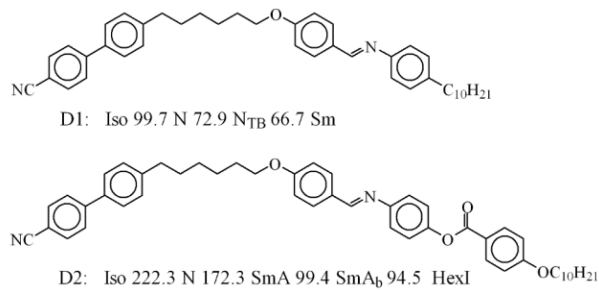

D1:  Iso 99.7 N 72.9 $N_{TB}$ 66.7 Sm

D2:  Iso 222.3 N 172.3 SmA 99.4 SmA$_b$ 94.5 HexI

**Fig. 1. Materials studied.** Molecular structures and phase transition temperatures (in deg. C, detected by calorimetric studies on cooling scan) for the compounds *D1* and *D2*.

Although for material *D1* calorimetric studies clearly revealed the enthalpy change associated with the transition between the $N_{TB}$ and smectic phases (Fig. S1), in optical studies this phase transition is not detectable, suggesting that in both phases there is a similar averaging of molecular position in space due to the formation of the helix. In cells with homeotropic anchoring condition, both phases are optically uniaxial. In homogenous cells, the *N* phase has a uniform texture while in the $N_{TB}$ phase a stripe texture (*27, 28*) develops and it persists into the smectic phase (Fig. 2, A and B).

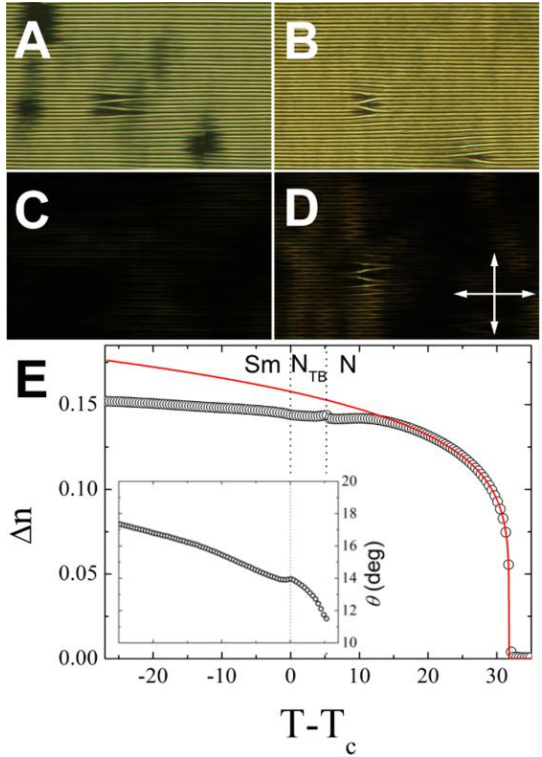

**Fig. 2. Optical studies.** Optical stripe texture in (**A**) the twist-bend nematic, $N_{TB}$, phase and (**B**) smectic phase observed for compound *D1* in a 3 μm thick planar cell. A nearly uniform texture in the (**C**) $N_{TB}$ phase and (**D**) smectic phase grown in sample kept for several minutes at a constant temperature. The polarizer directions are denoted by arrows. (**E**) The birefringence ($\Delta n$) of a uniform texture vs. temperature (T) measured on heating with red light (690 nm). The red line represents the fitting to the critical behavior $\Delta n = \Delta n_0 \left[(T - T_{Iso-N})/T_{Iso-N}\right]^\beta$ ($\Delta n_0 = 0.24$, $\beta = 0.18$, $T_{Iso-N} = 371.65$ K); the decrease of birefringence from the extrapolated $\Delta n$ value is due to the tilting of the molecules and formation of the heliconical structure. The inset: the conical tilt angle ($\theta$) vs. temperature, calculated from the changes in birefringence.

However, if the sample is kept in the $N_{TB}$ or smectic phase for several minutes at a constant temperature, the stripes start to disappear and a uniform texture is formed with the light extinction direction along the rubbing direction when viewed between crossed polarizers (Fig. 2, C and D). For a sample with a uniform texture, the birefringence measurements performed on heating show that there is only a minor change of birefringence upon the transition from the smectic to the $N_{TB}$ phase (Fig. 2E), which suggests that the conical angle (tilt) is similar in both phases. In the smectic phase, the tilt angle estimated from the birefringence measurements is ~15° (see details in the SM). On further heating, the birefringence increases on approaching the transition to the nematic phase, which is consistent with the decrease of the tilt angle in the heliconical phase (*29*).

The non-resonant x-ray measurements show that the layer spacing in the smectic phase is close to two molecular lengths (bilayer structure) (Fig. S2), which suggests a broken up-down symmetry of the molecular arrangement inside the smectic layers, often observed for cyano-biphenyl derivatives, due to the self-segregation effect of polar and non-polar end groups (*30*). Such a bilayer periodicity is also inherent to the local structure in the $N_{TB}$ and $N_{TB}$ phases, because the position of the diffuse x-ray diffraction signal in these phases coincides with the Bragg peak in the smectic phase (Fig. S2). In the high diffraction angle range in the smectic phase the diffuse signal, corresponding to 0.45 nm, evidences a liquid-like arrangement of the molecules in the smectic layers. In the resonant x-ray scattering studies, which reveal the periodic structures related to the orientational order, the signal associated with the helical structure develops at the $N - N_{TB}$ phase transition and moves continuously to higher values of the scattering wave vector magnitude ($q$) on reducing temperature (Fig. 3A), i.e., the pitch decreases with decreasing temperature.

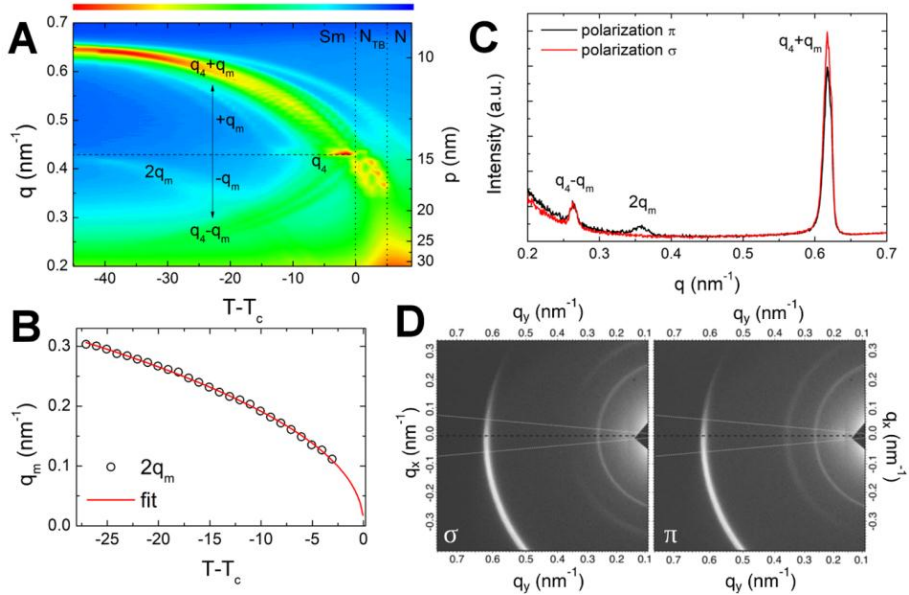

**Fig. 3. The resonant soft x-ray scattering for material D1.** (**A**) Temperature ($T$) evolution of the scattering wave vector magnitudes ($q$) and corresponding periodicities ($p$) of the RSoXS signals measured on heating, intensity of the signals is codded by colors, scale bar is given above. At the transition temperature ($T_c$) from the $N_{TB}$ to the smectic phase the signal locks at $q = q_4$, which corresponds to four smectic layer distances. This signal reflects the ideal clock-like helix in the smectic phase. For few degrees below the $N_{TB}$ – smectic phase transition the $q_4$ signal coexists with the signals $q_4 - q_m$ and $q_4 + q_m$, which are related to the distorted helical structure. The weak $2q_m$ signal is also visible. The range in which both structures, the distorted and ideal clock-like helix, coexists depends on the cooling/heating rate (see Fig. S2). (**B**) The temperature dependence of the $2q_m$ signal; the red line represents the fit to the critical dependence $2q_m = 2q_{m0}T_r^{0.45}$ with $2q_{m0} = 0.6$ nm$^{-1}$, $T_r = (T_c - T)/T_c$ being the reduced temperature. (**C**) Intensity of signals vs. wave vector magnitude $q$ obtained by integration of the 2D RSoXS patterns (**D**) registered for the $\pi$ and $\sigma$ polarization of the incident beam. Only the areas marked by white cones in (**D**) were analyzed. The $q_4 - q_m$ and $q_4 + q_m$ signals are independent of the polarization of the incident beam, whereas the $2q_m$ signal is polarization dependent.

At the $N_{TB}$ - smectic phase transition the RSoXS signal locks at the value $q_4$ corresponding to exactly 4 molecular distances. A departure from the resonant energy even by $2 - 3$ eV causes the signal to disappear, thus the signal associated with the 4-layer structure is purely resonant. Additionally, the $q_2$ signal related to the bi-layer periodicity is found (Fig. S3); although this signal is resonantly enhanced its intensity is very weak compared to the other detected signals. The continuous evolution of the structure at the $N_{TB}$ – smectic phase transition suggests that the helix in the smectic phase is made of molecules rotating on the tilt cone by exactly 90° between consecutive layers. The signal due to the 4-layer structure at the scattering wave vector magnitude $q_4$ persists over a few K below the $N_{TB}$ – smectic phase transition temperature, and this range is determined kinetically; it is wider at lower cooling rates (Fig. S3). On further cooling, a symmetric splitting of the $q_4$ signal into two signals: $q_4 + q_m$ and $q_4 - q_m$ is observed. The splitting increases with decreasing temperature (Fig. 3A) and both split signals have a resonant character. Their intensities are very different and for all the samples measured, the $q_4 + q_m$ signal is the more intense. Depending on the sample, the ratio of integrated intensities of the split signals is 15 to 50 throughout the whole temperature range. Simultaneously with the splitting of the $q_4$ signal, an additional, very weak signal is also detected at $2q_m$ in some temperature scans (Fig. 3A). The magnitude of the $2q_m$ scattering vector follows the critical temperature dependence (Fig. 3B), $2q_m = 2q_{m0}T_r^\beta$, where $T_r = (T - T_C)/T_C$ is the reduced temperature with $T_C$ being the $N_{TB}$ – smectic transition temperature. The fitting parameter $2q_{m0} = 0.6 \text{ nm}^{-1}$ is the theoretical magnitude of the modulation wave vector extrapolated to $T = 0$ K and $\beta = 0.45 \pm 0.05$ is the critical exponent associated with the phase transition. The $N_{TB}$ – smectic phase transition seems to be weakly first order with the mean field critical exponent. The RSoXS signals detected in the smectic phase have different anisotropy: both split signals $q_4 + q_m$ and $q_4 - q_m$ are nearly independent of the polarization ($\sigma$ or $\pi$) of the incident x-ray beam, whereas the intensity of the $2q_m$ signal is strongly sensitive to the polarization of the incident beam (Fig. 3, C and D).

What is the molecular structure behind the patterns observed using RSoXS in the smectic phase? For a structure with a repeating 4-layer basic unit, either with an ideal clock or distorted clock distribution of the molecules on a tilt cone or for a structure with molecules all-in-one-plane with consecutive synclinic and anticlinic interlayer interfaces, only a signal at $q_4$, (and its harmonics) should be observed (*31*). The splitting of the $q_4$ signal shows that additional modulation, with a longer periodicity, is superimposed over the 4-layer unit. The RSoXS pattern resembles closely that for the hierarchical structure of the ferrielectric $SmC^*_{FI2}$ phase, a tilted smectic subphase detected in chiral systems in a temperature range between the synclinic and anticlinic smectic $C$ phases (*32*). In the $SmC^*_{FI2}$ phase, molecules in 4 consecutive layers form a distorted helix, i.e., the azimuthal angle ($\delta$) between the projections of the long molecular axis to the smectic plane, in the pairs of the neighboring layers differs from $\pi/2$ (Fig. 4). Due to the chiral interactions the azimuthal angle in each layer increases by $\varepsilon$ with respect to the neighboring layer (*33*) and the whole 4-layer unit cell rotates forming a long wavelength 'optical' helix. However, it should be stressed that the materials on which we report in this study, are achiral. Therefore, even though the structure in the smectic phase of material $D1$ is similar to the $SmC^*_{FI2}$ phase, the interactions driving its formation must be different. Also, one should note that the modulation superimposed on the basic 4-layer unit is rather strong: deep in the smectic phase it is of the same order of magnitude as the basic modulation, i.e. on the nanometer scale.

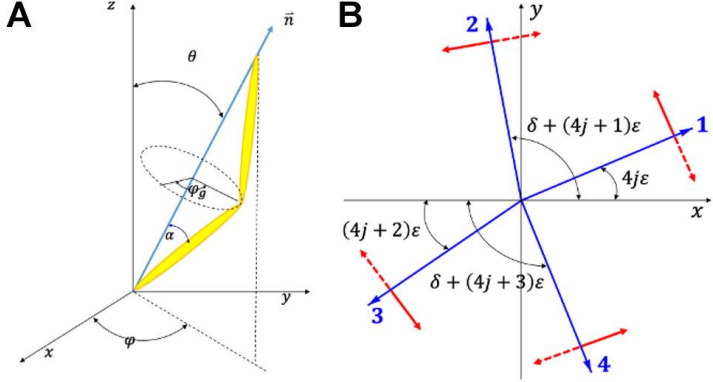

**Fig. 4. The ferri-like structure in material *D1*.** (**A**) The orientation of the long molecular axis (defined by the director $\vec{n}$) in the laboratory coordinate system is described by the tilt angle $\theta$ and the azimuthal angle $\varphi$. The apex angle of the molecule is $\pi - 2\alpha$ and the direction of the molecular tip with respect to the tilt plane (defined by the director $\vec{n}$ and the smectic layer normal, $\hat{z}$), is defined by the general tilt angle $\varphi_g$, which is zero, if the molecular tip is perpendicular to the tilt plane made of director $\vec{n}$ and the smectic layer normal, $\hat{z}$, in the direction of $\vec{n} \times \hat{z}$. The chirality at the molecular layer is defined by the direction of the molecular apex (sign of angle $\alpha$) with respect to the tilt plane (**B**) The projections of the director $\vec{n}$ on the smectic plane ($xy$-plane) in the $j$-th stack of four successive layers. The angle between the director projections in layers 1 and 2 and layers 3 and 4 is $\delta$, and the angle between the layers 1 and 3 and 2 and 4 is $\pi$. On this "basic" structure an additional rotation by an angle $\varepsilon$ is superimposed. The sign of $\delta$ and $\varepsilon$ define the helicity of 4-layer structure and helical structure superimposed on it, respectively. Thick red (solid or dashed) arrows denote the direction of the molecular tip when the tip is perpendicular to the tilt plane ($\varphi_g$ equals 0 or $\pi$). The direction of the molecular tip with respect to the tilt plane defines the layer chirality, in the model the layer chirality is preserved in consecutive layers. The two possible structures with positive and negative $\alpha$ were considered.

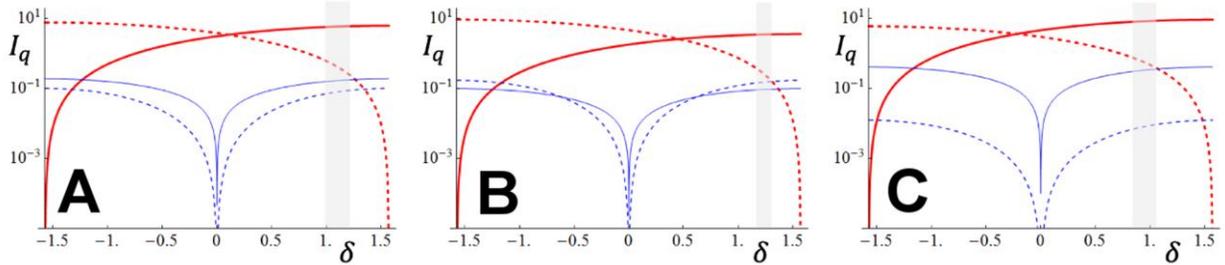

**Fig. 5. The scattered peak intensities for the model 4-layer structure.** The intensities ($I_q$) of the scattered light as a function of $\delta$ at $\theta = 0.25$ and $\varepsilon = 0.5$ for the peaks at $q = q_4 + q_m$ (red solid line: $\pi\sigma$ or $\sigma\pi$-polarizations), $q_4 - q_m$ (red dashed line: $\pi\sigma$ or $\sigma\pi$-polarizations), $q_2 + 2q_m$ (solid blue line: $\sigma\sigma$-polarizations) and $q_2 - 2q_m$ (dashed blue line: $\sigma\sigma$-polarizations) for (**A**) $\alpha = 0$, (**B**) $\alpha = 0.1$ and (**C**) $\alpha = -0.1$. There is a contribution to the $q_2 \pm 2q_m$ peak intensities also from the $\pi\pi$, $\pi\sigma$ or $\sigma\pi$-polarizations, but they are at least an order of magnitude lower than the $\sigma\sigma$ contribution. The gray shaded region shows the values of $\delta$ at which the ratio between the intensities of the $q_4 + q_m$ and $q_4 - q_m$ peaks is larger than 10 and at the same time both intensities are higher than the intensities of the $q_2 \pm 2q_m$ peaks, note that the regions are of different width for positive and negative values of $\alpha$.

To verify the proposed orientational order structure of the smectic phase, we constructed a model, formed of bent-core molecules; the orientation of their molecular long axis varies from one smectic layer to another in the manner similar as in the $SmC^*_{FI2}$ phase, as shown in Fig. 4. The scattering amplitude in the RSoXS experiment can be obtained by assuming that the tensor form factor describing the response of one resonant scatterer is proportional to the anisotropic part of the molecular polarizability tensor. The effect of all the carbon atoms that respond resonantly was considered by putting one resonant center in each arm of the molecule, with the polarizability being the largest along the direction along the molecular arm. Details of the calculation are given in SM, here we discuss the results. First, we considered the simplified case of the rod-like molecules ($\alpha = 0$). According to the model, only RSoXS signals $q_4 \pm q_m$, $q_2 \pm 2q_m$ and $2q_m$ should be observed; the full pitch signal at $q_m$ is forbidden. Fig. 5A gives the intensity of the peaks as a function of the distortion angle $\delta$ at fixed $\varepsilon$. The angle $\varepsilon$, which defines the additional modulation superimposed on the four layer unit can be deduced directly from the position of the $2q_m$ peak or alternatively from the splitting of the $q_4$ signal, and is $\varepsilon = 2\pi(2q_m)/q_0$, where $q_0$ is the wave vector associated with a single molecular layer thickness; for the material studied $q_0 = 2\pi d_0 = 1.8 \text{ nm}^{-1}$. The difference in the intensities of signals $I(q_4 + q_m)$ and $I(q_4 - q_m)$ approaches zero for an all-in-one plane model ($\delta = 0$ and $\varepsilon = 0$), while the difference is greatest for $\delta$ equal $\frac{\pi}{2}$ (ideal clock structure). Taking into account the experimentally detected ratio of the signal intensities (integrated signals) for the $q_4 + q_m$ and $q_4 - q_m$ peaks that is in the range from 15 to 50 and the fact that the intensity of the $q_2 \pm 2q_m$ peaks is below the detection limit, we find that $\delta \sim 60° - 70°$. The angle $\varepsilon$ near the $N_{TB}$ – smectic phase transition is a few degrees, and it increases to nearly 45° approximately 40 K below the transition. The extrapolated value $2q_{m0}$ obtained from the fitting of $2q_m$ to the power law corresponds to $\varepsilon \approx 70°$. It is worth noting that the sign of $\varepsilon$ with respect to that of $\delta$, uniquely defines the ratio of the split signal intensities. For all the measured samples, the intensity measured for the $q_4 + q_m$ peak is higher than for $q_4 - q_m$, which indicates that $\varepsilon$ and $\delta$ are coupled, therefore the handedness of the 'short' 4-layer 'distorted' helix (sign of $\delta$) determines the handedness of the 'long' helical modulations (sign of $\varepsilon$). The model correctly predicts also the polarization dependence of the diffraction signals (Fig. 3C). For the $\sigma$-polarized incident light (i.e. polarization in the direction perpendicular to the scattering plane), the $q_4 + q_m$ and $q_4 - q_m$ peaks are $\pi$-polarized (polarization in the scattering plane) and vice versa, which is in agreement with the experimentally observed lack of split peaks intensity changes upon changing the polarization of the incident beam for powder samples. The modulation peak, $2q_m$, behaves quite differently as it is strongly polarization dependent. For the $\sigma$-incident polarization, the peak has both $\sigma$ and $\pi$ components, the $\sigma$ component being much stronger. For the $\pi$-incident polarization, there is only a weak $\sigma$-polarized $2q_m$ peak.

Taking into account the nearly temperature independent value of the angle $\delta$ and a strongly temperature dependent value of the angle $\varepsilon$ the evolution of the helical structure can be reconstructed. The smectic phase just below the $N_{TB}$ phase has almost an ideal clock 4-layer structure ($\delta \sim \frac{\pi}{2}, \varepsilon \sim 0$). As the temperature is reduced the structure evolves towards an anticlinic one. The structures for a given $\varepsilon$ values and a 'virtual' structure obtained by increasing $\varepsilon$ to 90° are shown in Fig. 6 and in Movie S1.

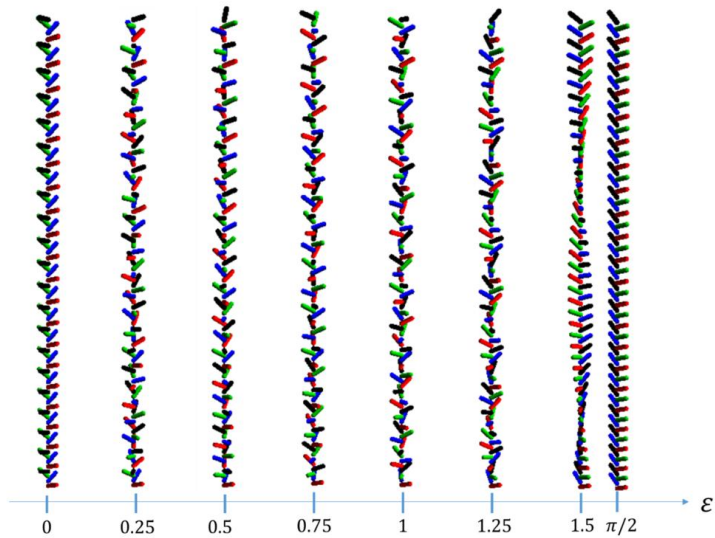

**Fig. 6. Temperature evolution of the 4-layer structure in material *D1*.** The structure as a function of $\varepsilon$ at $\delta = 1.2$ and $\theta = \pi/3$. The cone angle $\theta$ is chosen to be larger than measured value to better visualize the rotation of the molecules. The red, blue, green and black rods present the molecules in the successive 4 layers. Their position on the cone is given by the angles shown in Fig. 4. The values of $\varepsilon$ up to approximately $0.75$ are seen experimentally.

Although the model in which the molecules are represented by simple rods with the resonant atoms in the upper and lower part of the molecule appears to reproduce most of the experimental results, for the sake of completeness of theoretical approach we also considered a more general theoretical approach. The nonsymmetry of the molecular structure was introduced by placing two 'atoms' with different polarizabilities to represent the mesogenic cores of the dimeric nonsymmetric molecule. Such a nonsymmetry yields also the resonantly enhanced $q_2$ signal related to bilayers, as observed in the experiment. The bend of the molecular structure was also considered by a rotation of polarizability tensors representing two mesogenic cores in a molecule in the opposite directions by an angle $\alpha$ (see Fig. 4A). It should be noted that the bending of molecules, which are tilted with respect to the layer normal, introduces another level of structural chirality ('layer chirality'), as changing the tip direction with respect to the tilt plane results in the mirro-reflected structures (*34*). The 'layer' chirality is defined in the model by the sign of $\alpha$. The additional chirality level alters the relative signal intensities (Fig. 5, B and C). To account for the intensity of the $q_4 + q_m$ signal being 15-50 times higher than the intensity of the $q_4 - q_m$ peak and for the intensity of the $q_2 - 2q_m$ signal being below the detection limit, the molecular apex angle must be taken as being considerably larger than $120°$ ($\alpha \sim 10°$). This is in line with the observation that birefringence for bent dimeric molecules, in which the mesogenic cores are, on average, inclined by angle $\alpha$ from director, is not significantly smaller than for analogues monomeric compounds having mesogenic cores along director. Most probably, also the sign of 'layer' chirality is coupled to the handedness of the 4-layer basic unit and 'long-helix'; however, it cannot be unambiguously confirmed within the proposed model as it yields the proper ratio of the RSoXS signal intensities for both positive and negative values of $\alpha$ (Fig. 5, B and C).

Is the multihierarchical helical structure related only to materials with the $N_{TB}$ – smectic phase transition? By studying several bent dimer materials forming smectic phases, we found a similar structure also for material *D2*, which forms a tilted hexatic I (*HexI*) phase - a phase with a short positional ordering of molecules, but a long-range correlation of the local crystallographic axes direction (Bond Orientational Order, BOO) within the layers. The previous resonant x-ray studies for this homologue series revealed a simple 'clock' helix for the twist-bend smectic phase (*19*), which appears for the shorter homologues, $n = 7$ and 8. We studied a homologue with a longer terminal chain, $n = 10$, in which the hexatic phase is formed below the non-tilted biaxial orthogonal smectic phase ($SmA_b$). In the $SmA_b$ phase, the layer spacing is comparable to the molecular length; however, in the hexatic phase, non-resonant x-ray studies revealed a bilayer structure. Although the main diffraction signal corresponds to the molecular length, its weak subharmonic ($q_2$) related to the bilayer structure is also detected (Fig. S4) and it shows that either a weak self-segregation effect of polar and non-polar end groups occurs or the phase is of a general tilt type (*35*) in this temperature range. In the RSoXS measurements in the $HexI$ phase the bilayer signal $q_2$ and split peaks $q_2 - q_m$ and $q_2 + q_m$ are seen, the signal intensity ratio (integrated) of the latter two is ~100 over the whole temperature range, with the $q_2 + q_m$ signal being the stronger. The split signals disappear if the energy of the x-ray beam is off-resonant, proving that these signals are due to the modulation of molecular orientation. The splitting of the $q_2$ signal occurs simultaneously with the appearance of a low angle $2q_m$ signal corresponding to an approximately 55 nm periodicity. The $2q_m$ signal has the same azimuthal direction as the higher angle signals (Fig. 7). From the splitting of the $q_2$ peak and the position of the $2q_m$ peak we deduce that an additional helical rotation with $\varepsilon \sim 16°$ is superimposed over the basic bilayer structure. The asymmetry in the intensity of the $q_2 - q_m$ and $q_2 + q_m$ signals signifies that the structure is more complex than a simple analogue of the chiral anticlinic phase ($SmC_A^*$), with molecules in the consecutive layers on the opposite side of the tilt cone and the 'optical' helix is introduced by chiral interactions. For the $SmC_A^*$ structure the satellite signals are expected to have nearly equal intensities (*36, 37*). Interestingly, the periodicity ~55 nm (corresponding to $2q_m$ signal) can be also detected by AFM imaging (Fig. 8).

In the low angle range of the RSoXS pattern also a signal $q_f$ corresponding to a periodicity of 150 nm is also detected. For an aligned sample the azimuthal position of the $q_f$ signal is orthogonal to the $2q_m$ signal. The possible origin of this signal is the pitch of the mesoscopic helical filaments formed in the $HexI$ phase (Fig. 8) – the filaments resemble those found in the $B4$ phase (*38*). The structure of the $D2$ material is an interesting example of a multi-helical structure – the layered structure with a nanoscale helix formed by rotation of the molecules assemble into mesoscopic helical filaments.

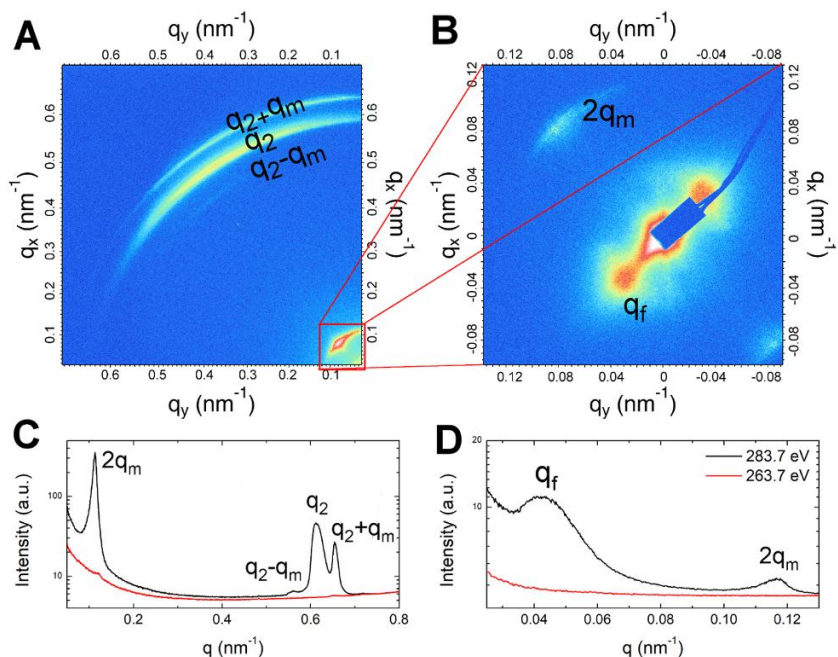

**Fig. 7. The resonant soft x-ray scattering for material *D2*.** 2D RSoXS patterns in the *HexI* phase registered at 80°C in a (**A**) wide and (**B**) small angle regimes. The signal intensity integrated over azimuthal angle in (**C**) wide and (**D**) small angle range; black line: intensity registered at carbon absorption K-edge, 283.7 eV; red line: intensity registered at off-resonant energy 263.7 eV.

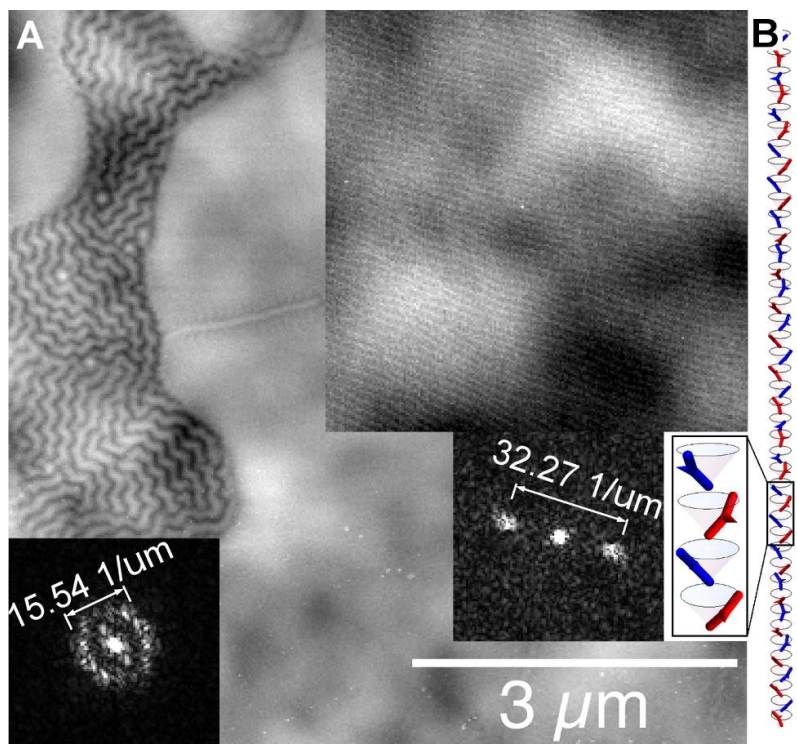

**Fig. 8. AFM studies of HexI phase of compound D2 and the structure model.** (**A**) Two morphologies coexist in samples of *HexI* phase of compound *D2*, most commonly observed are mesoscopic twisted filaments; however, in samples quickly quenched to room temperature also flat areas with parallel lines are observed; in upper right corner an enlarged flat area. FFT of the AFM image reveals the characteristic periodicities of ~150 nm and ~55 nm for filaments and lines, respectively. (**B**) Model of the helical HexI phase, the red and blue bent rods present the molecules in the neighboring layers. The arrows are added to better present the direction of the molecular tip. The structure is drawn for $\theta = 0.7$, $\alpha = -0.3$, $\varepsilon = 0.3$, $\delta = 0$. The cone angle is chosen larger than experimentally determined to better visualize the rotation of the molecules.

To analyze the RSoXS patterns and reveal the molecular orientation structure of the modulated smectic phase of compound *D2*, we modelled the bilayer unit in which the orientation of the molecules in the successive layers changes by an angle $\pi + \delta$, and the azimuthal position of the molecules is additionally modulated by an angle $\varepsilon$, giving a helix superimposed on the bilayer unit, as shown in Fig. 9A; details of the calculation are given in the SM. The model predicts that the intensity of the full pitch, $q_m$ peak, which is not observed experimentally, to be zero for an anticlinic structure. Therefore, for further analysis $\delta$ is set to zero. To account for the asymmetry in the intensity of the split signals $q_2 \pm q_m$ the molecular bend has to be considered by setting $\alpha \neq 0$, and in order to have the $q_2 + q_m$ peak more intense than the $q_2 - q_m$ peak, the $\varepsilon$ and $\alpha$ angles have to have opposite signs (Fig. 9B). Therefore, we conclude that as for the 4-layer structure, in the bi-layer structure the layer chirality is also coupled to the handedness of the 'long' helix. The experimentally determined ratio of the intensities between the $q_2 \pm q_m$ peaks that is ~ 100 can be obtained for $\alpha \approx -20°$ or $\alpha \approx -30°$. We have considered also a structure with the layer chirality switch between the neighboring layers ($\alpha$ of opposite sign in consecutive layers) (Fig. 9C). In this case the peaks $q_2 \pm q_m$ are of comparable intensities and the intensities of the $q_2 \pm 2q_m$ and $q_m$ peaks are not negligible. The bilayer structure as predicted by the combination of the model and experimental results is shown in Fig. 8B.

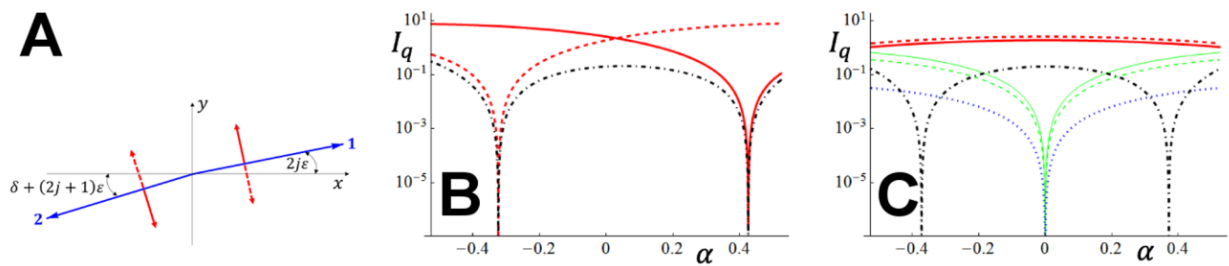

**Fig. 9 Model predictions for the bilayer structure.** (**A**) Molecules in the successive layers differ in the direction of the long molecular axis. The figure shows the projection of the director $\vec{n}$ on the smectic plane ($xy$-plane) in the $j$-th stack of two successive layers, on the nearly anticlinic structure of the director in the neighboring layers the additional rotation by angle $\varepsilon$ is superimposed. Thick red (solid or dashed) arrows denote the direction of the molecular tip when the tilt is perpendicular to the tilt plane ($\varphi_g$ equals 0 or $\pi$, see also Fig. 4), which defines the sign of the layer chirality. (**B** and **C**) The intensities ($I_q$) of the scattered light as a function of angle $\alpha$ defining the molecular bend, at $\theta = 0.4$, $\varepsilon = 0.3$ and $\delta = 0$ for the peaks at $q = q_2 + q_m$ (red solid line: $\pi\sigma$ or $\sigma\pi$-polarizations), $q_2 - q_m$ (red dashed line: $\pi\sigma$ or $\sigma\pi$-polarizations), $2q_m$ (dash-dotted black line: $\sigma\sigma$-polarizations), $q_2 + 2q_m$ (solid green line: $\sigma\sigma$-polarizations), $q_2 - 2q_m$ (dashed green line: $\sigma\sigma$-polarizations) and $q_m$ (dotted blue line: $\pi\sigma$ or $\sigma\pi$-polarizations) for the structure (**B**) with preserved layer chirality between the neighboring layers and (**C**) with the layer chirality switch

between the layers. For structure with the chirality switch we keep on graph both negative and positive part of $\alpha$ to emphasize the full symmetry of the graph with respect to this parameter.

In summary, both compounds studied formed smectic phases with multilevel chiral structures despite being composed of achiral molecules. At the lowest level, 'layer chirality' is defined by the direction of the molecular apex with respect to tilt plane. Chiral layers are stacked into 2- or 4-layer basic units for materials *D2* and *D1*, respectively; in the case of *D1* such a stacking is helical. The basic structure of both smectic phases is in addition helically modulated at a larger scale of several layers. Moreover, the material *D2* exhibits a chiral morphology in the *HexI* phase – it forms mesoscopic twisted filaments. Resonant X-ray scattering studies revealed that the sign of chirality at these different levels are coupled.

What is the cause of such an unusual complex structure for these smectic phases? Steric (*39*) and flexoelectric (*40*, *41*) interactions were suggested as a source of the short helix in the $N_{TB}$ phase, so we assume that these interactions are responsible also for the structure of the helical smectic phases. In the nematic phase these interactions lead to an 'ideal helix', the pitch of which decreases on reducing temperature as temperature-induced fluctuations decrease. The formation of smectic layers leads to a distortion of this ideal clock. As the temperature is lowered and competition between the twisting due to the flexoelectricity and the entropy-driven tendency for molecules to stay in one tilt plane increases, the structure continuously evolves from nearly an ideal 4-layer clock to nearly anticlinic bilayer. For the achiral dimers studied here, this is a continuous evolution while for the chiral systems a set of discontinuous phase transitions between structures with 4-, 3- and 2-layer basic units was observed (*1*). For material *D2* for which interlayer interactions are strong system forms double helix with pitch nearly temperature independent in broad temperature range.

**Acknowledgments**
**Funding:** M.S., D.P. and N.V. acknowledge the support of the National Science Centre (Poland) under the grant no. 2016/22/A/ST5/00319. E.G. acknowledges the founding from Foundation for Polish Science through the Sabbatical Fellowships Program. N.V. acknowledges the support of the Slovenian Research Agency (ARRS), through the research core funding No. P1-0055. RW gratefully acknowledges the Carnegie Trust for the Universities of Scotland for funding the award of a PhD scholarship. The beamline 11.0.1.2 at the Advanced Light Source at the Lawrence Berkeley National Laboratory are supported by the Director of the Office of Science, Office of Basic Energy Sciences, of the U.S. Department of Energy under Contract No. DE-AC02-05CH11231.

# Multi-level Chirality in Liquid Crystals Formed by Achiral Molecules

Mirosław Salamończyk[1,2], Nataša Vaupotič[3,4*], Damian Pociecha[1], Rebecca Walker[5], John M. D. Storey[5], Corrie T. Imrie[5], Cheng Wang[2], Chenhui Zhu[2], Ewa Gorecka[1*]

[1]University of Warsaw, Faculty of Chemistry, Żwirki i Wigury 101, 02-089 Warszawa, Poland.

[2]Lawrence Berkeley National Laboratory, Advanced Light Source, 1 Cyclotron Rd, Berkeley, CA 94720, USA.

[3]Department of Physics, Faculty of Natural Sciences and Mathematics, University of Maribor, Koroška 160, 2000 Maribor, Slovenia.

[4]Jozef Stefan Institute, Jamova 39, 1000 Ljubljana, Slovenia.

[5]Department of Chemistry, King's College, University of Aberdeen, Aberdeen AB24 3UE, UK.

*Correspondence to: natasa.vaupotic@um.si (N.V.), gorecka@chem.uw.edu.pl (E.G.).

**This PDF file includes:**

> Experimental Methods
> Supplementary Text
> Figs. S1 to S8
> Caption for Movie S1

**Experimental Methods**

The resonant x-ray experiments were performed on the soft x-ray scattering beam line (11.0.1.2) at the Advanced Light Source of Lawrence Berkeley National Laboratory. The x-ray beam was tuned to the K-edge of carbon absorption with the energy ~280 eV. The x-ray beam with a cross-section of $300 \times 200$ μm was linearly polarized, with the polarization direction that can be continuously changed from the horizontal to vertical. Samples with thickness lower than 1 μm were placed between two 100-nm-thick Si3N4 slides. The scattering intensity was recorded using the Princeton PI-MTE CCD detector, cooled to $-45°C$, having a pixel size of 27 μm, with an adjustable distance from the sample. The detector was translated off axis to enable a recording of the diffracted x-ray intensity. The adjustable position of the detector allowed to cover a broad range of $q$ vectors, corresponding to periodicities from approximately $5.0 - 500$ nm.

Optical studies were performed by using the Zeiss Imager A2m polarizing microscope equipped with Linkam heating stage. Samples were observed in glass cells with various thickness: from 1.8 to 10 µm.

Calorimetric studies were performed with TA DSC Q200 calorimeter, samples of mass from 1 to 3 mg were sealed in aluminum pans and kept in nitrogen atmosphere during measurement, both heating and cooling scans were performed with a rate of $5 - 10$ K/min.

The broad angle X-ray diffractograms were obtained with the Bruker D8 GADDS system (CuKα line, Goebel mirror, point beam collimator, Vantec2000 area detector). Samples were prepared as droplets on a heated surface. The temperature dependence of the layer thickness was determined from the small-angle X-ray diffraction experiments performed with the Bruker D8 Discover system (CuKα line, Goebel mirror, Anton Paar DCS350 heating stage, scintillation counter) working in the reflection mode. Homeotropically aligned samples were used, prepared as a thin film on a silicon reflectionless wafer.

The birefringence was calculated from the optical retardation at red light ($\lambda = 690$ nm). The retardation was measured with a setup based on a photoelastic modulator (PEM-90, Hinds) working at a modulation frequency $f = 50$ kHz; as a light source a halogen lamp (Hamamatsu LC8) was used equipped with a narrow band pass filter (633 nm). The signal from a photodiode 113 (FLC Electronics PIN-20) was deconvoluted by a lock114 in amplifier (EG&G 7265) into $1f$ and $2f$ components to yield the retardation induced by the sample. Knowing the sample thickness, the retardation was recalculated into optical birefringence. For birefringence measurements the 3 µm – thick cells were used with a planar alignment layer. Because the system allows to measure retardation only up to $\lambda/2$, the absolute value of retardation was determined by comparing the results obtained for the samples with different thickness, 1.8 and 5 µm and assuming a continuous evolution of retardation near the phase transition to the isotropic phase. The changes of birefringence in the nematic phase were fitted by assuming a critical dependence $\Delta n = \Delta n_0 [(T - T_C)/T_C]^\beta$, where $\Delta n_0$, $T_C$ and $\beta$ were free parameters. The conical tilt angle was estimated from a decrease of birefringence in the $N_{TB}$ and smectic phase from the extrapolated value of $\Delta n$ found in the nematic phase as $\Delta n_{tilt} = 1/2 \, \Delta n \, (3 \cos^2 \theta - 1)$ (29).

The AFM images were taken with the Bruker Dimension Icon microscope, working in the tapping mode at the liquid crystalline-air surface. Cantilevers with a low spring constant 0.4 N/m were used, the resonant frequency was in a range of $70 - 80$ kHz, a typical scan frequency was 1 Hz. Samples for the AFM imaging were prepared on microscopy cover glass at an elevated temperature and quenched to room temperature.

**Supplementary Text**

Theoretical modelling
In modelling the response in the resonant x-ray scattering (42, 43) which provides the information on the orientational structure of molecules, we model a bent-core molecule as having two resonant dipoles with a uniaxial polarizability. The tensor form factor $\underline{F}_{ei}$ in the eigensystem of the dipole, where the local $z$-axis is along the direction in which polarizability is largest, is proportional to the anisotropic part of the polarizability tensor (*31*):

$$\underline{F}_{ei} = f_0 \begin{pmatrix} 1 & 0 & 0 \\ 0 & 1 & 0 \\ 0 & 0 & -2 \end{pmatrix},$$

where $f_0$ is a parameter that depends on the scattering strength. When measuring at the carbon K-edge, several atoms in the molecule respond, and $f_0$ is expected to be much larger that in case when there is only one resonant atom, like, for example, Sulphur, built into the molecule. The tensor form factor in the laboratory system is obtained by a set of the following rotations of $\underline{F}_{ei}$ (for angle definitions see Fig. S5 (A)):
- rotation by angle $\alpha$ (lower arm of the molecule) or $-\alpha$ (upper arm of the molecule) around the local axis $y$;
- rotation by angle $\varphi_g$ around the $z$-axis;
- rotation by angle $\theta$ around the $x$-axis;
- rotation by angle $\varphi$ around the $z$-axis.

The tensor form factor in the laboratory system is thus:
$$\underline{F} = \underline{R}_\varphi \underline{R}_\theta \underline{R}_{\varphi_g} \underline{R}_\alpha \underline{F}_{ei} \underline{R}_\alpha^T \underline{R}_{\varphi_g}^T \underline{R}_\theta^T \underline{R}_\varphi^T \tag{1}$$

where $\underline{R}_\theta$, $\underline{R}_\varphi$, $\underline{R}_{\varphi_g}$ and $\underline{R}_\alpha$ are rotation matrices:

$$\underline{R}_\theta = \begin{pmatrix} 1 & 0 & 0 \\ 0 & \cos\theta & \sin\theta \\ 0 & -\sin\theta & \cos\theta \end{pmatrix},$$

$$\underline{R}_\varphi = \begin{pmatrix} \cos\varphi & \sin\varphi & 0 \\ -\sin\varphi & \cos\varphi & 0 \\ 0 & 0 & 1 \end{pmatrix},$$

$$\underline{R}_{\varphi_g} = \begin{pmatrix} \cos\varphi_g & \sin\varphi_g & 0 \\ -\sin\varphi_g & \cos\varphi_g & 0 \\ 0 & 0 & 1 \end{pmatrix}$$

and

$$\underline{R}_\alpha = \begin{pmatrix} \cos\alpha & 0 & \sin\alpha \\ 0 & 1 & 0 \\ -\sin\alpha & 0 & \cos\alpha \end{pmatrix}.$$

Once we have the tensor form factor, the scattering amplitude tensor ($\underline{A}$) is obtained by the summation of the form factor over several (theoretically infinite) number of layers. The tensor elements reduce to delta functions (these define the magnitudes of the scattering vectors) multiplied by some factor, the norm of which is related to the scattering intensity at a given polarization of the scattered waves in dependence of the incident polarization of waves.

The intensity of the scattered light is calculated for the polarization of the incident and scattered light being either in the scattering plane ($\pi$-polarization) or perpendicular to it ($\sigma$-polarization). There are thus four possibilities to check for each RSoXS peak.

The unit vectors defining the direction of the $\sigma$ and $\pi$ polarization can be deduced from Fig. S5(B):

$$\vec{\pi}_{in} = \left(\sin\left(\frac{\theta_{sc}}{2}\right), 0, \cos\left(\frac{\theta_{sc}}{2}\right)\right),$$

$$\vec{\pi}_{sc} = \left(-\sin\left(\frac{\theta_{sc}}{2}\right), 0, \cos\left(\frac{\theta_{sc}}{2}\right)\right),$$

$$\vec{\sigma}_{in} = (0,1,0),$$

$$\vec{\sigma}_{sc} = (0,1,0).$$

The scattering angle ($\theta_{sc}$) is related to the scattering vector magnitude ($q$):

$$\sin\left(\frac{\theta_{sc}}{2}\right) = \frac{q}{2k_0},$$

where $k_0$ is the magnitude of the wave vector of the incident/scattered light. The intensities of the peaks $I_{nm}$, where $n = \sigma, \pi$ denotes the polarization of the scattered light and $m = \sigma, \pi$ polarization of the incident light, are (*42*):

$$I_{\sigma\sigma} = \left|\vec{\sigma}_{sc} \cdot \underline{A} \cdot \vec{\sigma}_{in}\right|^2,$$

$$I_{\pi\sigma} = \left|\vec{\pi}_{sc} \cdot \underline{A} \cdot \vec{\sigma}_{in}\right|^2,$$

$$I_{\sigma\pi} = \left|\vec{\sigma}_{sc} \cdot \underline{A} \cdot \vec{\pi}_{in}\right|^2,$$

$$I_{\pi\pi} = \left|\vec{\pi}_{sc} \cdot \underline{A} \cdot \vec{\pi}_{in}\right|^2.$$

Below we study the scattering intensities and polarization of the scattered peaks for the four-layer and bilayer modulated smectic structures.

1) Four – layer structure

The tensor form factor of the ferri-like 4-layer structure is obtained by summing up the form factors (eq. (1)) by taking into the account the orientation of the long molecular axis in each smectic layer (Fig. S6) and by considering the phase difference due to the scattering at different layers, the thickness of each being $d_0$. Within each layer we put two resonant scatterers: one in the center of the lower and one in the center of the upper arm of the molecule. We assume, that, in general, they can have slightly different polarizability (different $f_0$) and that the neighboring arms of molecules from two different layers have the same polarizability. The form factors for these two types of dipoles are denoted by $\underline{F_1}$ and $\underline{F_2}$ (i.e., $f_0 = f_{01}$ for one dipole and $f_0 = f_{02}$ for the other dipole). The tilt angle $\theta$ and the apex angle $\pi - 2\alpha$ are the same for all layers. The tilt of the molecular arm ($\alpha$) is opposite in the lower and upper arm of the molecule. First, we assume that there is no general tilt, which means that the rotation matrix $\underline{R}_{\varphi_g}$ is a unit matrix. The scattering amplitude tensor ($\underline{A}_4$) is obtained as a combination of the form factor due to a four layer structure and structure factor due to the 4-layer repeating unit:

$$\underline{A}_4 = \sum_{j=1}^{N} \big(F_1(\theta, \alpha, 4j\varepsilon) + F_2(\theta, -\alpha, 4j\varepsilon)e^{iqd_0/2} + F_2(\theta, \alpha, \delta + (4j+1)\varepsilon)e^{iqd_0}$$
$$+ F_1(\theta, -\alpha, \delta + (4j+1)\varepsilon)e^{3iqd_0/2}$$
$$+ F_1(\theta, \alpha, \pi + (4j+2)\varepsilon)e^{i2qd_0}$$
$$+ F_2(\theta, -\alpha, \pi + (4j+2)\varepsilon)e^{i5qd_0/2}$$
$$+ F_2(\theta, \alpha, \pi + \delta + (4j+3)\varepsilon)e^{i3qd_0}$$
$$+ F_1(\theta, -\alpha, \pi + \delta + (4j+3)\varepsilon)e^{i7qd_0/2}\big)e^{4ijqd_0} \quad , \qquad (2)$$

where $N$ is the number of layers on which x-rays are scattered. The magnitude of the scattering vector is chosen as

$$q = \frac{2\pi}{d_0} h \quad , \qquad (3)$$

where $h$ is the Miller index. Because the basic periodicity is $4d_0$ (if $\varepsilon = 0$), one expects interference peaks at multiples of $1/4$. When $\varepsilon \neq 0$, these peaks split.

In eq. (2), the summation over a very large (infinite) number of layers leads to delta functions and the tensor elements are different from zero only for $\pm 8\varepsilon + 8\pi h = 2\pi m$ and $\pm 4\varepsilon + 8\pi h = 2\pi m$, where $m$ is an integer, which means:

$$h = \frac{m}{4} \pm \frac{\varepsilon}{\pi}$$

and

$$h = \frac{m}{4} \pm \frac{\varepsilon}{2\pi} \quad .$$

We are interested in the integer $m$ being $0, 1$ or $2$. The magnitudes of the scattering vectors related to the allowed peaks are denoted by $q_2 \pm q_m$, $q_2 \pm 2q_m$, $q_4 \pm q_m$, $q_4 \pm 2q_m$, $q_m$ and $2q_m$, where $q_4 = \pi/(2d_0)$, $q_2 = \pi/d_0$ and $q_m = \varepsilon/d_0$.

Let us write the scattering amplitude tensor $\underline{A}_4$ in a general form:

$$\underline{A}_4 = \begin{pmatrix} f_{11} & f_{12} & f_{13} \\ f_{21} & f_{22} & f_{23} \\ f_{31} & f_{32} & f_{33} \end{pmatrix} \quad .$$

The summation over $j$ in eq.(2) gives that the tensor elements $f_{11}, f_{22}, f_{12}$ and $f_{21}$ are different from zero at $h = \frac{m}{4} \pm \frac{\varepsilon}{\pi}$, $f_{11}$ and $f_{22}$ also at $h = \frac{m}{4}$, $f_{33}$ only at $h = \frac{m}{4}$, while $f_{13}, f_{31}, f_{23}$ and $f_{32}$ are different from zero only at $h = \frac{m}{4} \pm \frac{\varepsilon}{2\pi}$.

We present the results for the intensities of the peaks assuming symmetric molecules ($f_{01} = f_{02}$). The ratio of the intensities ($I_{4-}/I_{4+}$) of the peaks at $q_4 \pm q_m$ as a function of angle $\delta$ is given in Fig. S7 (A). For the presentation of the results we have chosen $\theta = 0.25$ (the experimentally estimated value) and $\varepsilon = 0.5$ (the value in the middle of the temperature range of the modulated smectic phase). Experimentally, the intensity of the $q_4 + q_m$ peak is by an order of magnitude higher than the intensity of the $q_4 - q_m$ peak. We have added a line to the plot at $I_{4-}/I_{4+} = 0.05$, which corresponds to the intensity of the $q_4 + q_m$ peak being 20-times higher than the intensity of the $q_4 - q_m$ peak. We see, that the value of $\delta$ at which the intensity ratio crosses this line depends on the value of $\alpha$ and it is different for structures with opposite layer chirality. We see, that we obtain the proper ratio of the $q_4 \pm q_m$ intensities if the angle $\delta$ is larger than 1. This value seems reasonable, because the modulated smectic structure is formed below the twist-bend nematic phase in which one observes a strong resonant peak due to ideal helix with pitch length approaching the four – layer periodicity. We note, that the value of $\delta$ at which the intensity ratio curve crosses the $I_{4-}/I_{4+} = 0.05$ line slightly depends also on the value of $\varepsilon$. In presentation of further results we use $\delta = 1.2$.

Figures S7 B and C give the intensities of the peaks $2q_m$, $q_4 \pm q_m$ and $q_2 \pm q_m$. The peaks $q_m$, $q_4 \pm 2q_m$ and $q_2 \pm q_m$ are zero in this model, also if asymmetry is considered. If a general tilt is considered, as well, then the intensity of these peaks becomes different from zero. One special case of the general tilt is $\varphi_g = \pi$ in every second layer. This means that there is a layer chirality switch from one layer to another. Because there are no experimentally observed $q_m$, $q_4 \pm 2q_m$ and $q_2 \pm q_m$ peaks, we conclude, that the general tilt, if present, is very small and that there is no chirality switch from one layer to another. From Figs. S7 B and C we see, that the peak intensities vary with $\alpha$ and $\theta$. Because no splitting of the $q_2$ peak is observed experimentally, we expect the intensities of the $q_2 \pm 2q_m$ peaks to be much lower than the intensities of the $q_4 \pm q_m$ peaks. They are lower if $\alpha$ is small, so we conclude, that the apex angle of molecules in material *D1* that forms a 4-layer structure is close to $\pi$. In addition, we can expect that $\alpha$ is positive in order to have the $q_4 - q_m$ stronger than the $q_2 \pm 2q_m$ peaks (see Fig. S7(B). The intensities of the $q_2 \pm 2q_m$ peaks reduce also by reducing the angle $\delta$. The change in the modulation angle $\varepsilon$ does not have a significant effect on the intensities (and the ratio between them).

Figure 6 in the main text shows the orientation of molecules as a function of $\varepsilon$ in a stack of several layers in order to visualize the temperature development of the helical structure. We see, that the system tries to unwind into a simple anticlinic structure, although this situation is not experimentally reachable. A video clip showing the development of the structure by a continuous increase of $\varepsilon$ is available among the supporting material.

2) Bilayer structure

To model the bilayer structure, we model the director positions on the cone as shown in Fig. S6 (B). The scattering amplitude tensor ($\underline{A}_2$) is obtained by following the procedure given in the construction of the 4-layer structure (see the text above eq. (2)):

In eq. (4), the summation over a very large (infinite) number of layers leads to delta functions and the tensor elements are different from zero only for $\pm 4\varepsilon + 4\pi h = 2\pi m$ and $\pm 2\varepsilon + 4\pi h = 2\pi m$, where $m$ is an integer, which means:
$$h = \frac{m}{2} \pm \frac{\varepsilon}{\pi}$$

and

$$h = \frac{m}{2} \pm \frac{\varepsilon}{2\pi},$$

where we are interested in the integer $m$ being 0 or 1. The magnitudes of the scattering vectors related to the allowed peaks are denoted by $q_2 \pm q_m$, $q_2 \pm 2q_m$, $q_m$ and $2q_m$, where $q_2 = \pi/d_0$ and $q_m = \varepsilon/d_0$.

Fig. S8 shows the intensities of the $q_2 \pm q_m$, $q_2 \pm 2q_m$, $q_m$ and $2q_m$ peaks at experimental values of $\theta$ and $\varepsilon$ ($\theta = 0.4$, $\varepsilon = 0.3$) as a function of $\delta$. For $\alpha$ we take $\alpha = \pm\pi/6$ (the apex angle is 120 deg). Because experimentally no $q_2 \pm 2q_m$ and $q_m$ peaks are observed, from graphs in Fig. S8 we conclude that $\delta \approx 0$, i.e. the structure is anticlinic with the helical modulation superimposed over it. We also see that the layer chirality determines the wave vector of the peak with the highest intensity. If $\alpha < 0$ this is the peak with $q = q_2 + q_m$, while if $\alpha > 0$, the peak with $q = q_2 - q_m$ has the highest intensity. In experiment, the peak with $q = q_2 + q_m$ has higher intensity than the peak with $q = q_2 - q_m$, so we conclude that $\alpha < 0$.

**Supplementary Figures**

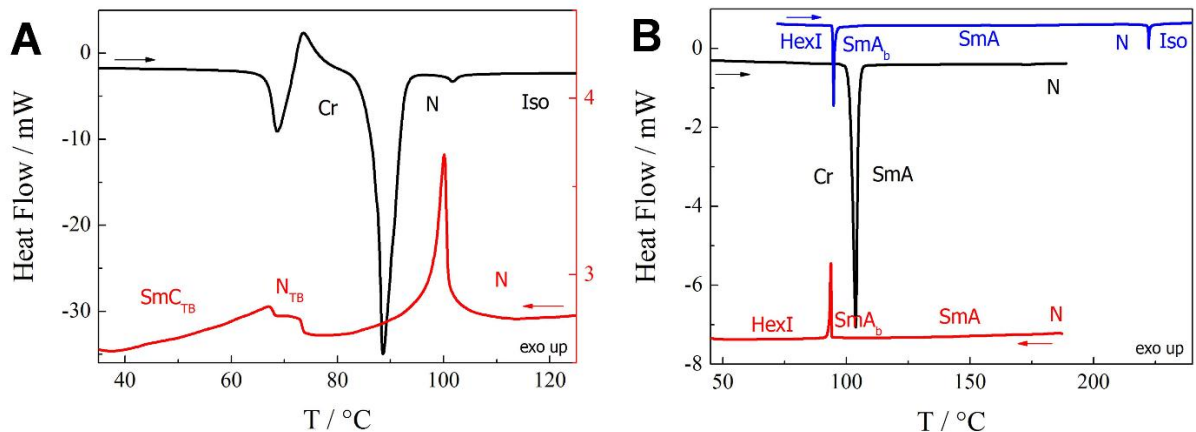

**Fig. S1.**
DSC thermograms for heating and cooling scans for compounds (**A**) *D1* and (**B**) *D2*. Blue line in **B** shows second heating run, which was started from supercooled $HexI$ phase in order to record monotropic $HexI - SmA_b$ phase transition on heating.

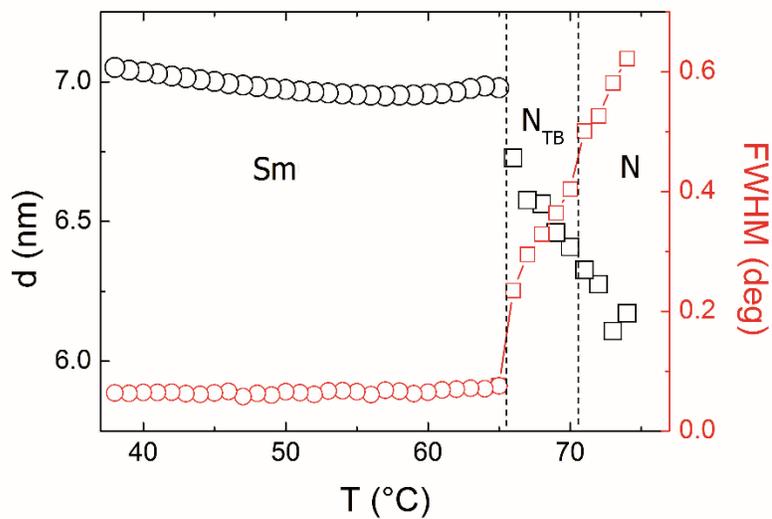

**Fig. S2.**
Layer thickness ($d$) vs. temperature ($T$) (black) and the width of the signal ($FWHM$) vs. temperature (red) for material *D1*.

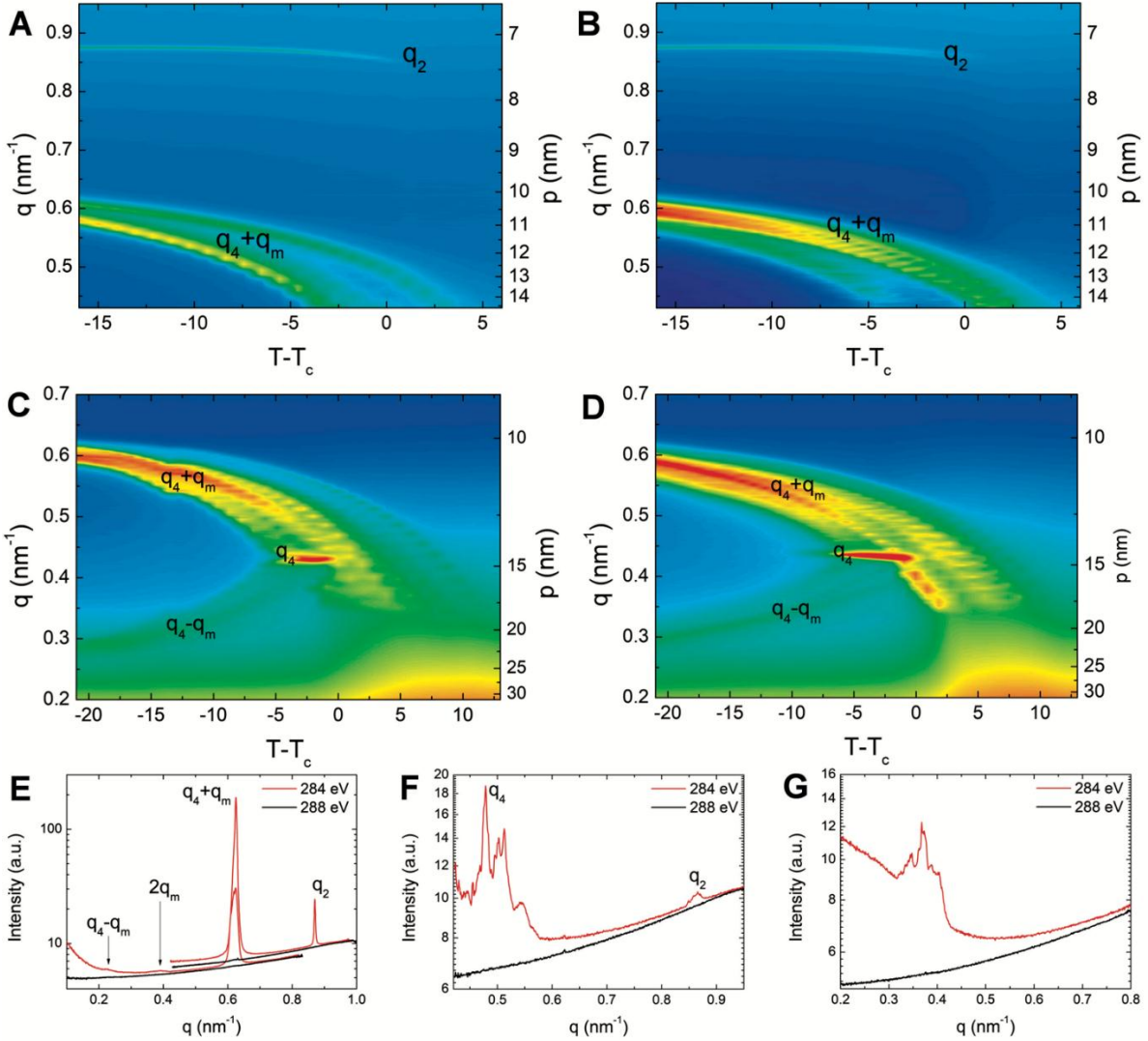

**Fig. S3.**
The consecutive RSoXS scans for material *D1* obtained at (**A, C**) heating and (**B, D**) cooling covering different scattering vector magnitude ($q$) ranges, $p$ is the corresponding modulation pitch. The intensity vs. $q$ registered at the resonant energy 284 eV and out of resonance at 288 eV in the smectic phase (**E**) 20 K and (**F**) 3 K below the $N_{TB}$ – smectic phase transition, and in (**G**) the $N_{TB}$ phase 5 K above the $N_{TB}$ – smectic phase transition. The scattering vectors due to the 4-layer and bilayer periodicities are $q_4$ and $q_2$, respectively, $q_m$ is the modulation wave vector.

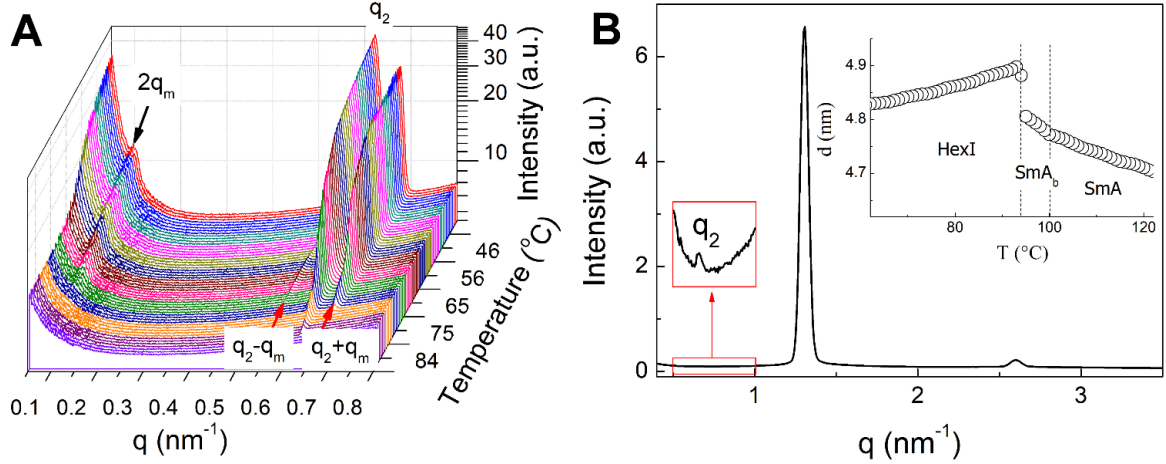

**Fig. S4.**
(**A**) Intensity vs. scattering wave vector magnitude ($q$) and temperature ($T$) for material $D2$ in the hexatic phase ($HexI$) obtained from the RSoXS measurements. The $q_2$ peak corresponding to the bilayer periodicity, a half-pitch modulation peak at $q = 2q_m$ and two peaks at $q = q_2 \pm q_m$ are observed in the $HexI$ phase. (**B**) Intensity vs. $q$ for material $D2$ in the hexatic phase obtained from the nonresonant x-ray scattering; the inset: the layer thickness ($d$) evolution with temperature.

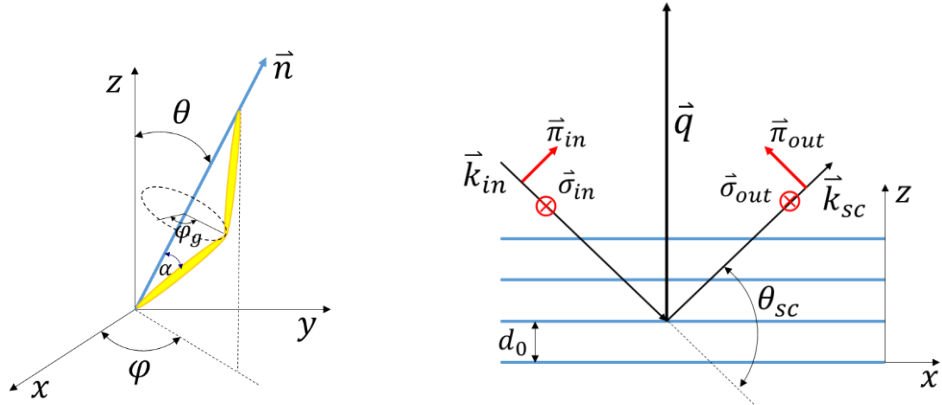

**Fig. S5.** (**A**) The orientation of the long molecular axis (defined by the director $\vec{n}$) in the laboratory coordinate system is defined by the tilt angle $\theta$ and the azimuthal angle $\varphi$. The apex angle of the molecule is $\pi - 2\alpha$ and the direction of the molecular tip with respect to the tilt plane, defined by the director $\vec{n}$ and the smectic layer normal (direction $z$), is defined by the general tilt angle $\varphi_g$, which is zero, if the molecular tilt is in the direction perpendicular to the tilt plane in the direction of $\vec{n} \times \hat{z}$. (**B**) The scattering geometry; $\vec{k}_{in}$ and $\vec{k}_{out}$ are the wave vectors of the incident and scattered wave, respectively, $\vec{q}$ is the scattering vector; $\vec{\sigma}_{in,out}$ and $\vec{\pi}_{in,out}$ are polarizations of the incident and scattered light, $\theta_{sc}$ is the scattering angle and $d_0$ is the smectic layer thickness.

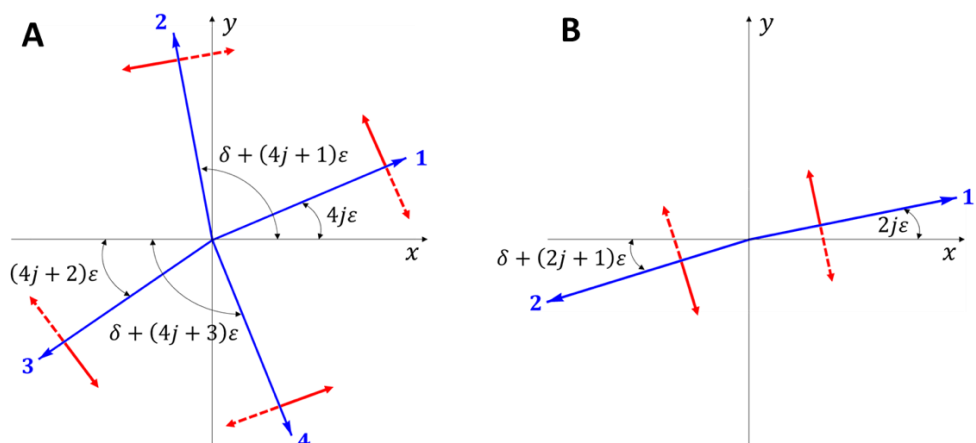

**Fig. S6.** (**A**) In the four layer structure the molecules in the successive layers differ in the direction of the long molecular axis direction. The figure shows the projection of the director $\vec{n}$ on the smectic plane ($xy$-plane) in the $j$-th stack of four successive layers. The structure is ferri-like: the angle between the director projections in layers 1 and 2 and layers 3 and 4 is $\delta$, and the angle between the layers 1 and 3 and 2 and 4 is $\pi$. To this "basic" ferri-like structure an additional rotation by an angle $\varepsilon$ is superimposed. (**B**) The bilayer structure. To an almost anticlinic structure of the director in the neighboring layers the additional rotation by angle $\varepsilon$ is superimposed. Both figures: Thick red (solid an dashed) arrows denote the direction of the molecular tip when the tilt is perpendicular to the tilt plane ($\varphi_g$ equals 0 or $\pi$).

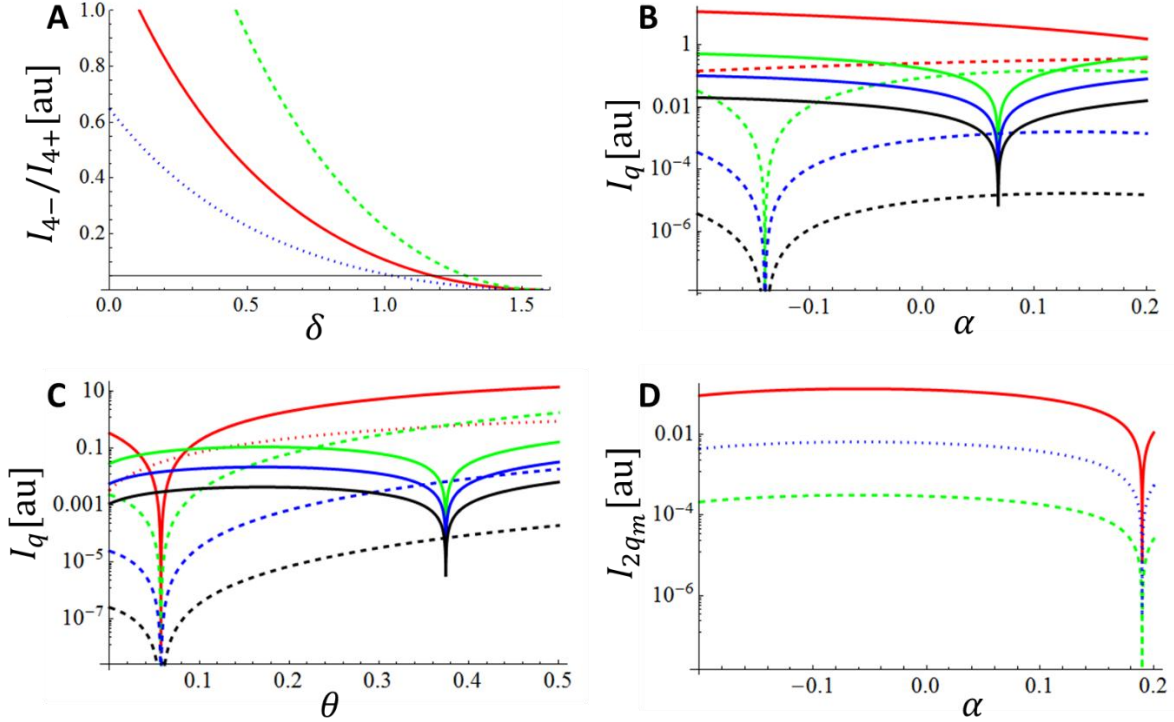

**Fig. S7.** (**A**) Ratio between the intensities of the $q_4 - q_m$ ($I_{4-}$) and $q_4 + q_m$ ($I_{4+}$) peaks as a function of $\delta$ at $\theta = 0.25$ and $\varepsilon = 0.5$ for $\alpha = 0$ (red solid line), $\alpha = -0.1$ (blue dotted line) and $\alpha = 0.1$ (green dashed line). (**B**) The intensities ($I_q$) of the scattered light as a function of $\alpha$ at $\theta = 0.25$, $\delta = 1.2$ and $\varepsilon = 0.5$ for the peaks at $q = q_4 + q_m$ (red solid line: $\pi\sigma$ or $\sigma\pi$-polarizations), $q_4 - q_m$ (red dashed line: $\pi\sigma$ or $\sigma\pi$-polarizations), $q_2 + 2q_m$ (solid green line: $\sigma\sigma$-polarizations; blue solid line: $\pi\sigma$ or $\sigma\pi$-polarizations; black solid line: $\pi\pi$-polarization) and $q_2 - 2q_m$ (dashed green line: $\sigma\sigma$-polarizations; blue dashed line: $\pi\sigma$ or $\sigma\pi$-polarizations; black dashed line: $\pi\pi$-polarizations). (**C**) The intensities of the scattered light as a function of $\theta$ at $\alpha = 0.1$, $\delta = 1.2$ and $\varepsilon = 0.5$ for the peaks at $q_4 + q_m$ (red solid line: $\pi\sigma$ or $\sigma\pi$-polarizations), $q_4 - q_m$ (red dashed line: $\pi\sigma$ or $\sigma\pi$-polarizations), $q_2 + 2q_m$ (solid green line: $\sigma\sigma$-polarizations; blue solid line: $\pi\sigma$ or $\sigma\pi$-polarizations; black solid line: $\pi\pi$-polarization) and $q_2 - 2q_m$ (dashed green line: $\sigma\sigma$-polarizations; blue dashed line: $\pi\sigma$ or $\sigma\pi$-polarizations; black dashed line: $\pi\pi$-polarizations). (**D**) The intensities ($I_{2q_m}$) of the scattered light as a function of $\alpha$ at $\theta = 0.25$, $\delta = 1.2$ and $\varepsilon = 0.5$ for the peaks at $2q_m$; red solid line: $\sigma\sigma$-polarizations; blue dotted line: $\pi\sigma$ or $\sigma\pi$-polarizations; green dashed line: $\pi\pi$-polarizations.

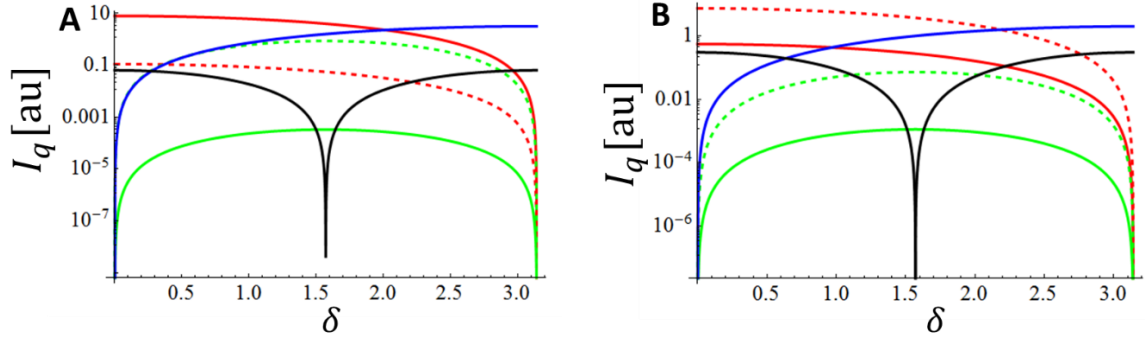

**Fig. S8.** The intensities ($I_q$) of the scattered light as a function of $\delta$ at $\theta = 0.4$, $\varepsilon = 0.3$ and (**A**) $\alpha = -\pi/6$ and (**B**) $\alpha = \pi/6$ for the peaks with the scattering wave vector equal to $q_2 + q_m$ (solid red line), $q_2 - q_m$ (dashed red line), $q_2 + 2q_m$ (solid green line), $q_2 - 2q_m$ (dashed green line), $q_m$ (solid blue line) and $2q_m$ (solid black line). The $q_2 \pm q_m$ and $q_m$ peaks are $\sigma$-polarized for the $\pi$-polarized incident light and vice versa. The intensities of the $q_2 \pm 2q_m$ and $2q_m$ peaks are the highest for the $\sigma$-polarized incident and scattered light and only these intensities are shown. For other combinations of polarizations the scattered intensity is by at least one order of magnitude lower.